\documentclass[a4paper, 10pt, conference]{ieeeconf}      % Use this line for a4
\IEEEoverridecommandlockouts                              % This command is only
\usepackage[utf8]{inputenc}
\usepackage[T1]{fontenc}
\usepackage{graphicx}
\usepackage{listings}
\usepackage{graphicx}
\usepackage{subfig}
\usepackage[ruled,vlined]{algorithm2e}
\usepackage{array}
\usepackage{tabularx}
\title{\LARGE \bf
MapVisual: A Visualization Tool for Memory Access Patterns
}

\author{ \parbox{5 in}{\centering Pavlos Aimoniotis, Maria-Rafaela Gkeka and Nikolaos Bellas\\
        \small \textit{Department of Electrical and Computer Engineering\\
        University of Thessaly \\}
        {\tt\small \{paimoniotis, margkeka, nbellas\}@uth.gr}}
}

\author{Pavlos Aimoniotis, Maria Rafaela Gkeka and Nikolaos Bellas$^{*}$% <-this % stops a space
\thanks{$^{1}$\{paimoniotis, margkeka, nbellas\}@inf.uth.gr}%
}

\begin{document}

\maketitle
\thispagestyle{empty}
\pagestyle{empty}

%%%%%%%%%%%%%%%%%%%%%%%%%%%%%%%%%%%%%%%%%%%%%%%%%%%%%%%%%%%%%%%%%%%%%%%%%%%%%%%%
\begin{abstract}

 Memory bandwidth is strongly correlated to the complexity of the memory access pattern of a running application. To improve memory performance of applications with irregular and/or unpredictable memory patterns, we need tools to analyze these patterns during application development. In this work, we present a software tool for the analysis and visualization of memory access patterns. We perform memory tracing and profiling, we do data processing and filtering, and we use visualization algorithms to produce three dimensional graphs that describe the patterns both in space and in time. Finally, we evaluate our toolflow on a variety of applications. 

\end{abstract}

%%%%%%%%%%%%%%%%%%%%%%%%%%%%%%%%%%%%%%%%%%%%%%%%%%%%%%%%%%%%%%%%%%%%%%%%%%%%%%%%
\section{INTRODUCTION}

Memory has traditionally been a major performance bottleneck in computing systems. Complex data structures and dynamic memory allocations may result in complex access patterns scattered all over the memory space. Additionally, the memory has a predefined size that does not allow an unlimited number of programs to run concurrently. For many years, the "Memory Wall" has been a major problem referring to the growing speed gap between CPU and DRAM.
%\vspace{-10pt}
%\begin{figure}[ht!]
%\centering
%    \includegraphics[width = 0.4 \textwidth]{img/memwall.jpg}
%    \caption{The growing CPU-DRAM speed gap~\cite{machanick2002approaches}}
%    \label{fig:cpu_dram}
%\end{figure}
%\vspace{-5pt}
\cite{grun2001apex}, \cite{boncz1999database}, \cite{jang2010exploiting} have mentioned that one way to improve memory performance is by optimizing memory access patterns (MAP). The difference between an L1 cache-hit, and a full miss resulting in main-memory access, is about 50ns to 100ns. If algorithms make random walks, they are less likely to benefit from the hardware support that hides this latency.

Using code analysis and visualization, our contribution provides new insights to assist code developers to better understand memory access patterns.  There have been several tools for code and memory profiling, such as \textit{Intel Advisor} and \textit{gprof}. Although these tools provide excellent profiling, they lack visualization. The MapVisual implementation flow is described briefly in the following paragraph.

First, (1) we use a memory trace file and we analyze it through the Gleipnir tool. Second, (2) we parse the Gleipnir output file and (3) we go through a data filtering process to produce analytics. Finally, (4) we perform visualization techniques on the output of step (3).

\section{BACKGROUND}
%\vspace{-2pt}
\subsection{Valgrind Framework}

Valgrind is an open-source dynamic binary instrumentation (DBI) framework used for memory debugging, memory leak detection and profiling. Valgrind
is a Virtual Machine that uses Just-In-Time compilation techniques to translate the assembly code into a processor-agnostic Intermediate Representation (IR). A Valgrind tool (such as Lackey) can process the IR before Valgrind translates the IR back into machine code for host-based execution.

\vspace{-3pt}
\subsection{Gleipnir}
\label{gleipnirintro}
Gleipnir is a memory analysis tool that maps an application’s source code variables to generated data traces~\cite{janjusic2013gleipnir}. It is an extension of Valgrind’s Lackey. The trace file generated by Lackey is purely a data and instruction trace which includes only load and store instructions and their corresponding data address. Gleipnir extends this information with the thread id, originating function, scope, data structure and the corresponding element, or a single scalar variable name. Gleipnir produces the following information \textit{Operation}, \textit{Address}, \textit{Memory\_Size}, \textit{Thread\_Id}, \textit{Scope}, \textit{Function}, \textit{Variable\_Info}.

Gleipnir traces and produces an output line for every load, store and modify in memory. This tracing can lead to a significant run time overhead. MapVisual optimizes the Gleipnir tool specifically for our purposes by filtering the amount of information it produces. 

%\begin{figure}[h!]
%\centering
%    \includegraphics[width=\columnwidth]{img/gleipnir.png}
%    \caption{Figure from \cite{janjusic2011international}: Listing 1 and Listing 2 show an example of trace information collected by Gleipnir. Listing 1 shows a very simple program that contains a few data structures and a single function. Listing 2 shows a segment of the trace file generated by Gleipnir for program code in Listing 1. The format of the output generated by Gleipnir is straightforward.
%}
%    \label{fig:gleipnir}
%\end{figure}

\vspace{-3pt}
\begin{figure*}[hbt!]
    \centering
    \setkeys{Gin}{width=0.3\textwidth}
\subfloat[Blocked Matrix Multiplication (BMM) Complete Memory Access Pattern Visualization
          \label{fig:cmap}]{\includegraphics[width=0.26\textwidth]{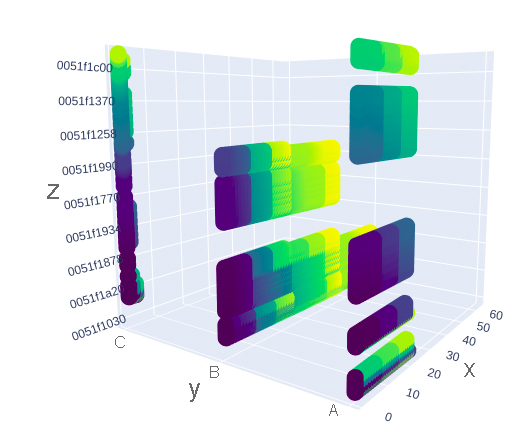}}
    \hfill
\subfloat[Blocked MM C = A*\textbf{B}, Array B 2D Visualization
          \label{fig:2dmap}]{\includegraphics[width=0.26\textwidth]{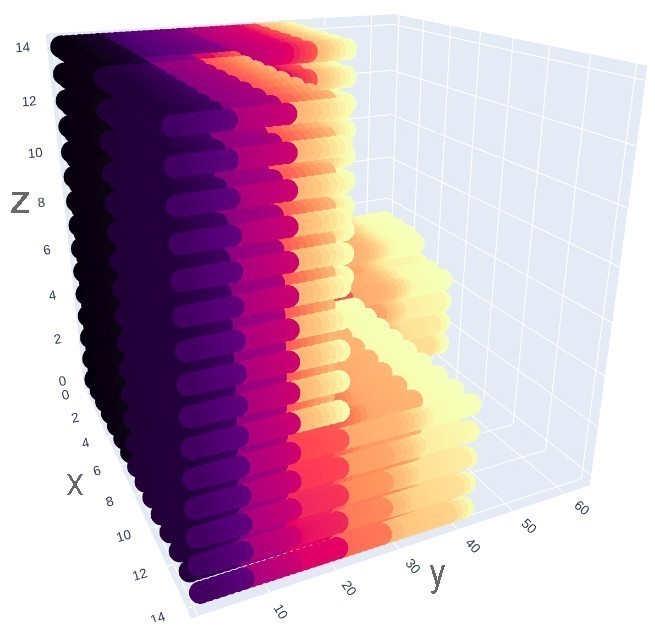}}
    \hfill
\subfloat[3D Array Accessed Sequentially Visualization
          \label{fig:3dmap}]{\includegraphics[width=0.31\textwidth]{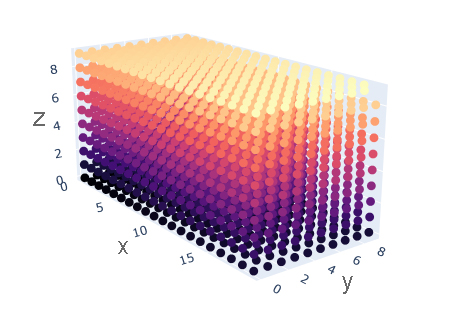}}
\vspace{-3pt}
\caption{Memory Access Patterns Visualization for three Applications}
\label{fig:visualizations}
\vspace{-10pt}
\end{figure*}

\section{MapVisual Analysis}
\subsection{Memory Tracing} 
\label{memorytracing}
Memory tracing is crucial for the performance of the tool. We run Gleipnir to generate the information on the allocated memory, the variable name and address starting point. We used the following two flags:
\vspace{-2pt}
\begin{lstlisting}
--read-var-info=yes 
--read-debug=yes 
\end{lstlisting}
\vspace{-2pt}
However, this creates major bottlenecks regarding time and space performance. We reduce the generated trace file $25\%$ on average, with bigger applications tending to be around $75\%$. We observed a peak reduction of $83\%$ on one of our benchmarks, achieving significant timing performance speedup. 

\vspace{-3pt}
\subsection{Data Structure}
\label{Datastruct}

We parse the output file of Gleipnir in order to minimize the amount of information produced~\cite{uththesis}. We create record instances for every valid address from the trace file and a Look-Up Table (LUT) for every unique record. For every duplicate of an address, we create a list with the \textit{(1) Operation Type}, representing stores, loads and modifies, and the \textit{(2) LUT ID}, representing the position of that access in the LUT to avoid performance overhead of storing the same data for every address. 
%\vspace{-12pt}
%\begin{figure}[ht!]
%\centering
%    \includegraphics[width=7cm]{img/datafilter.pdf}
%    \caption{Data Organization}
%    \label{fig:datafilt}
%\vspace{-8pt}  
%\end{figure}

%\begin{figure}[hbtp]
%    \label{gleipnirout}
%    \centering
%    \begin{lstlisting}[language=Mathematica]
%            X START 0:32372 at 0
%            X THREAD_CREATE 0:1 
%            X MMAP 004050000 12288
%            X 1 MALLOC 0051f1030 40 tangledaccess_c_12 0
%            X 1 MALLOC 0051f1080 40 tangledaccess_c_13 0
%            S 1ffefff468 8 1 S   
%            L 1ffefff468 8 1 S   
%            L 1ffefff490 8 1 S LV arr 
%            L 1ffefff49c 4 1 S LV i 
%            S 0051f1030 4 1 H H-0 tangledaccess_c_12.0 
%            L 1ffefff49c 4 1 S LV i 
%            S 0051f1080 4 1 H H-0 tangledaccess_c_13.0
%    \end{lstlisting}
%    \caption{Gleipnir Sample Output}
%\end{figure}

%We also process separate information produced in the beginning of the trace file, regarding memory allocations, which will be helpful to identify array names and elements position.

\vspace{-3pt}
\subsection{Data Filtering}
\label{datafilter}

The goal is to provide analytics prior to visualization. We determine for individual memory addresses \textit{Memory Usage}, \textit{Appearances} and \textit{Variable Name}.

\textit{Memory usage} refers to statistics collected about operations per address. \textit{Appearances} refers to how many times each address has been accessed during the execution of the program. \textit{Variable Name} refers to the name of an array. The process of extracting the name is mandatory for heap allocated elements, as Gleipnir does not provide such information straightforward.

We memory traced and stored data sequentially, as a consequence tool structures represent real-time accessibility. A timeline of the execution is generated using this particular information.

\vspace{-3pt}
\subsection{Visualization}
\label{visualization}
We produce three types of visualization. 
\begin{enumerate}
\item{\textit{Complete memory access pattern}, refers to the whole execution of the program and is the standard visualization of the tool. This 3D visualization demonstrates for each variable the addresses and their usage.}
\item{\textit{Two-dimension array visualization}, refers to 2D arrays. It reconstructs an array in a three-dimensional world. The third dimension represents information about memory usage.}
\item{\textit{Three-dimension array visualization}, refers to 3D arrays. It reconstructs an array in a three-dimensional world, where the axes show the position of each element.}
\end{enumerate}

To visualize, we process the data we created of ordered information in \ref{datafilter} and allocated information in \ref{Datastruct}. Visualizations have different coloring, a heat map that refers to the execution timeline of the program. Also, multiple features are supported, such as zoom in, zoom out, and mouse-over information. 
%For the visualizations, we used Plotly library.

\vspace{-2pt}
\section{EXPERIMENTAL RESULTS}

Figure~\ref{fig:visualizations} depicts the visualizations of three example benchmarks. Each one corresponds to the types mentioned in \ref{visualization}. The color heat map shows how far in the past a specific address was accessed: a darker color indicates an older access, whereas a brighter color indicates a more recent memory access.

Figure \ref{fig:cmap} shows the Complete MAP of a Blocked Matrix Multiplication (BMM) $C = A * B$, axes $z$ and $y$ represent addresses and variables respectively. The $x-axis$ (depth) represents the number of accesses per address and the color represents the occurrence time of an access during the execution of the program. Using this information we can observe the blocks accessed during the multiplication. Figure \ref{fig:2dmap} shows a reconstructed 2D array of BMM and how this array has been accessed. In this visualization, axes represent rows $(z-axis)$ and columns $(x-axis)$ of an array, and number of appearances per array element $(y-axis)$. In particular, this visualization represents array B of $C = A * B$ BMM, and shows the memory access pattern in this particular array. From the different coloring and number of accesses per element we can determine the different pattern and how code utilized this array. Element (0,0) that represents the start of the array is located at the bottom left, due to Plotly library implementation. Finally, Figure \ref{fig:3dmap} shows a reconstruction of a 3D array that has been accessed sequentially. Axes $(z, x, y)$ represent the coordinates of 3D array elements. This particular array has been accessed first in depth $y-axis$, then towards columns $x-axis$ and finally towards rows $z-axis$. This is inferred from the slow color change along the depth axis which means those elements are accessed near in time, the slightly more frequent color change in columns, and finally the abrupt color change in rows, where we conclude that there has been a long interval between the two accesses. 

We also evaluated our work in various other applications, such as Raycast Algorithm and Mandelbrot Set.
\vspace{-5pt}
\section{CONCLUSIONS}

In this work, we presented a software for the visualization of Memory Access Patterns. We demonstrated the memory tracing process, the implementation of the tool and ideas on visualizing memory access patterns. Although further optimizations can be applied, we provide a strong tool that can be crucial on MAP analysis.

\bibliographystyle{IEEEtran}
\bibliography{bibliography}

% Generated by IEEEtran.bst, version: 1.14 (2015/08/26)
\begin{thebibliography}{1}
\providecommand{\url}[1]{#1}
\csname url@samestyle\endcsname
\providecommand{\newblock}{\relax}
\providecommand{\bibinfo}[2]{#2}
\providecommand{\BIBentrySTDinterwordspacing}{\spaceskip=0pt\relax}
\providecommand{\BIBentryALTinterwordstretchfactor}{4}
\providecommand{\BIBentryALTinterwordspacing}{\spaceskip=\fontdimen2\font plus
\BIBentryALTinterwordstretchfactor\fontdimen3\font minus
  \fontdimen4\font\relax}
\providecommand{\BIBforeignlanguage}[2]{{%
\expandafter\ifx\csname l@#1\endcsname\relax
\typeout{** WARNING: IEEEtran.bst: No hyphenation pattern has been}%
\typeout{** loaded for the language `#1'. Using the pattern for}%
\typeout{** the default language instead.}%
\else
\language=\csname l@#1\endcsname
\fi
#2}}
\providecommand{\BIBdecl}{\relax}
\BIBdecl

\bibitem{grun2001apex}
P.~Grun, N.~Dutt, N.~Dutt, N.~Dutt, and A.~Nicolau, ``Apex: access pattern
  based memory architecture exploration,'' in \emph{Proceedings of the 14th
  international symposium on Systems synthesis}.\hskip 1em plus 0.5em minus
  0.4em\relax ACM, 2001, pp. 25--32.

\bibitem{boncz1999database}
P.~A. Boncz, S.~Manegold, M.~L. Kersten \emph{et~al.}, ``Database architecture
  optimized for the new bottleneck: Memory access,'' in \emph{VLDB}, vol.~99,
  1999, pp. 54--65.

\bibitem{jang2010exploiting}
B.~Jang, D.~Schaa, P.~Mistry, and D.~Kaeli, ``Exploiting memory access patterns
  to improve memory performance in data-parallel architectures,'' \emph{IEEE
  Transactions on Parallel and Distributed Systems}, vol.~22, no.~1, pp.
  105--118, 2010.

\bibitem{janjusic2013gleipnir}
T.~Janjusic and K.~Kavi, ``Gleipnir: A memory profiling and tracing tool,''
  \emph{ACM SIGARCH Computer Architecture News}, vol.~41, no.~4, pp. 8--12,
  2013.

\bibitem{uththesis}
\BIBentryALTinterwordspacing
C.~Ntogkas, ``Design and implementation of a tool for memory access pattern
  visualization,'' \emph{University of Thessaly, Diploma Thesis}, 2017.
  [Online]. Available:
  \url{https://www.e-ce.uth.gr/wp-content/uploads/formidable/59/Ntogkas\_Christos.pdf}
\BIBentrySTDinterwordspacing

\end{thebibliography}

\end{document}